\documentclass[10pt,aps,twocolumn,prd,superscriptaddress,showpacs,nofootinbib,noshowkeys,floatfix,preprintnumbers]{revtex4}
\usepackage{graphics,graphicx}
\usepackage{amsmath, amssymb}
\usepackage{multirow}
\usepackage{longtable}
\usepackage{color}

\usepackage[normalem]{ulem}  

\begin{document}
\title{Probing Deconfinement with Polyakov Loop Susceptibilities}
\date{\today}
\author{Pok Man Lo}
\affiliation{GSI, Helmholzzentrum f\"{u}r Schwerionenforschung,
Planckstr. 1, D-64291 Darmstadt, Germany}
\author{Bengt Friman}
\affiliation{GSI, Helmholzzentrum f\"{u}r Schwerionenforschung,
Planckstr. 1, D-64291 Darmstadt, Germany}
\author{Olaf Kaczmarek}
\affiliation{Fakult\"at f\"ur Physik, Universit\"at Bielefeld, 33615 Bielefeld, Germany}
\author{Krzysztof Redlich}
\affiliation{Institute of Theoretical Physics, University of Wroclaw,
PL-50204 Wroc\l aw, Poland}
\affiliation{Extreme Matter Institute EMMI, GSI,
Planckstr. 1, D-64291 Darmstadt, Germany}
\author{Chihiro Sasaki}
\affiliation{Frankfurt Institute for Advanced Studies, D-60438 Frankfurt am Main, Germany}

\begin{abstract}
{The susceptibilities of the real and imaginary parts, as well as of the modulus of the
Polyakov loop, are computed in $SU(3)$ lattice gauge theory. We show that  the ratios of these susceptibilities are
excellent probes of the deconfinement transition, independent of the renormalization of the Polyakov loop
and only weakly dependent on the system size.
The ratios are almost temperature independent above and below the transition and exhibit a discontinuity
at the transition temperature. This characteristic behavior can be understood in terms of the global $\mathcal{Z}_3$  symmetry
of the Yang-Mills Lagrangian and the general properties of the Polyakov loop probability distribution.
}
\end{abstract}
\maketitle

Systems described by a pure $SU(N_c)$ gauge theory in (d+1)-dimensions
undergo a phase transition at finite temperature.
Owing to the conjectured universality with  d-dimensional $\mathcal{Z}_{N_c}$ spin systems \cite{sy1},
this transition is of  general interest.

For $N_c=3$,  the transition is first order and is
characterized by spontaneous breaking of the global $\mathcal{Z}_{3}$
center symmetry of the Yang-Mills Lagrangian \cite{sy2,Boyd:1995zg,su31,ref1,Borsanyi:2012ve}.

The Polyakov loop,  which is linked to the free energy of a static quark immersed in
a  hot gluonic medium~\cite{McLerran:1980pk,general},  can be used to define an order
parameter of the deconfinement transition.
At low temperatures its thermal expectation value 
vanishes,  implying color confinement, while
at high temperatures it is non-zero, resulting in a finite
energy of a static  quark and consequently deconfinement of color.
While the basic thermodynamic functions of  $SU(3)$ pure gauge theory are
well established within the lattice approach  
\cite{sy2,Boyd:1995zg,su31,ref1,Borsanyi:2012ve,ref2,ref3,ref4,ref5}, the situation is less satisfactory
for the renormalized Polyakov loop and, in particular, for the corresponding susceptibilities.

In a pure $SU(3)$ gauge theory, the temperature of the confinement-deconfinement
transition is uniquely defined by the discontinuity of the order parameter, since the transition is first order.
More generally, for systems where the transition is continuous, e.g. QCD, the transition temperature
is identified by a maximum of the fluctuations, quantified e.g. by one of the Polyakov loop susceptibilities.
For $N_{c}\geq 3$, the Polyakov loop
operator is complex valued. Correspondingly, one can define
susceptibilities  of the real and imaginary parts as well as of the modulus of the Polyakov loop.

In a Yang-Mills theory, formulated on the  lattice,
the ultraviolet divergence of the bare quark-antiquark free energy implies
that, in the continuum limit, the bare Polyakov loop vanishes at any temperature.
Thus, in order to obtain a physically meaningful continuum limit,  the Polyakov loop must
be renormalized~\cite{Kaczmarek:2002mc,gavai}.
The renormalization of gluon correlation functions in general, and the Polyakov
loop susceptibility in particular, are still subject to uncertainties.


In this paper we bypass these ambiguities by considering
the ratios of Polyakov loop susceptibilities. In particular,
we focus on their  properties  near the deconfinement transition. To
this end, we compute the temperature de\-pen\-den\-ce of the
Polyakov loop susceptibilities within SU(3) lattice gauge
theory on different sized lattices and
examine the relevance of susceptibility ratios
as probes of the deconfinement transition.

We argue that these characteristics are naturally understood in terms of the
global $\mathcal{Z}_3$ symmetry  and general
properties of the Polyakov loop probability distribution.
Moreover, they are independent of the renormalization of the Polyakov loop and
depend only weakly on the volume. This implies that the susceptibility ratios are excellent observables for
identifying the confinement-deconfinement phase transition in
$SU(3)$ pure gauge theory.

\vskip 0.2cm
\noindent
{{\it The Polyakov loop  susceptibilities on the lattice.}}
On a $N_\sigma^3\times N_\tau$ lattice,  the Polyakov loop is defined as the trace of the product over temporal gauge links,
\begin{eqnarray}
        L_{\vec x}^{\rm bare} ={\frac{1}{N_c}} Tr \prod_{\tau=1}^{N_\tau} U_{(\vec x,\tau),4}\, ,\\
L^{\rm bare} = \frac{1}{N_\sigma^3}\sum_{\vec x} L_{\vec x}^{\rm bare}\, .
\end{eqnarray}
Due to the  $\mathcal{Z}_3$ symmetry of the pure gauge action, this quantity vanishes,  when
averaged over all gauge field configurations. Furthermore the Polyakov loop  is
strongly $N_\tau$ dependent and must be renormalized.

\begin{figure}[t]
\includegraphics[width=3.355in]{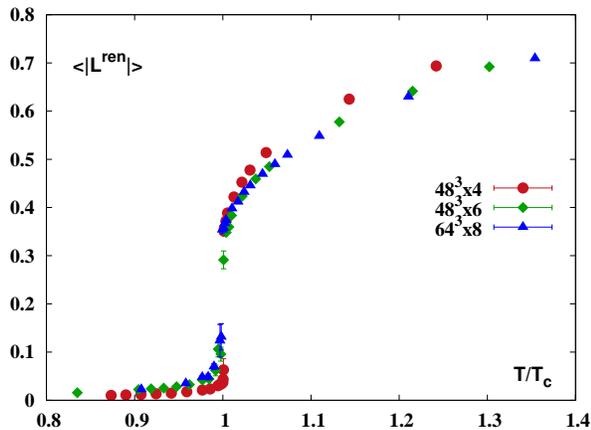}
\caption{
The modulus of renormalized Polyakov loop $\langle \vert L^{\rm ren} \vert \rangle$ obtained in $SU(3)$ lattice gauge theory.
 }
    \label{PL}
\end{figure}

These problems are avoided by considering the renormalized
Polyakov loop \cite{Kaczmarek:2002mc},
\begin{eqnarray}\label{r1}
 L^{\rm ren} = \left(Z(g^2)\right)^{N_\tau} L^{\rm bare}
\end{eqnarray}
and introducing the ensemble average of the modulus thereof, $
\langle \vert L^{\rm ren} \vert \rangle
$.  The latter is well defined in the continuum and  thermodynamic limits  and is an order parameter for the
spontaneous breaking of the   $\mathcal{Z}_3$ center symmetry. The lattice gauge theory result for $ \langle \vert L^{\rm ren} \vert \rangle$, as a function of temperature, is shown in Fig.~\ref{PL}.

As noted above, the location of the  phase transition is correlated with a maximum (or divergence) of
the fluctuations of the order parameter. For the confinement-deconfinement transition,
these fluctuations are reflected in the renormalized Polyakov loop susceptibility\footnote{In the following
we deal only with the renormalized Polyakov loop, and hence drop the superscript on $L^{\rm ren}$.}
\begin{eqnarray}\label{eq:chiA}
T^3 \chi_A =& \frac{N_\sigma^3}{N_\tau^3} \left( \langle \vert L \vert^2 \rangle - \langle
   \vert L \vert \rangle^2\right).
\end{eqnarray}

In the $SU(3)$ gauge theory, the Polyakov loop operator is complex. Consequently,
in addition  to $\chi_A$, one can also explore independent fluctuations of the real
and imaginary parts of the Polyakov loop.  Taking the $\mathcal{Z}_{3}$ symmetry into account,
we define a longitudinal and a transverse susceptibility\footnote{There is no mixing between longitudinal and transverse susceptibilities.}
\begin{eqnarray}
T^3 \chi_{L} =& \frac{N_\sigma^3}{N_\tau^3} \left[ \langle  (L_{L})^2 \rangle - \langle
    L_{L} \rangle^2\right].\label{eq:chiR}\\
T^3 \chi_{T} =& \frac{N_\sigma^3}{N_\tau^3} \left[ \langle  (L_{T})^2 \rangle - \langle
    L_{T} \rangle^2\right],\label{eq:chiI}
    \end{eqnarray}
where
$L_{L}={\rm Re}(\tilde{L})$
and
$L_{T}={\rm Im}(\tilde{L})$.
Here we have introduced the $\mathcal{Z}_{3}$ transformed Polyakov loop, $\tilde{L}=L\,e^{2\pi n i/3}$,
with $n=0,\pm1$. The phase of the transformation is chosen
such that for $T>T_{c}$ the expectation value of the transformed Polyakov loop, $\langle\tilde{L}\rangle$, is real.
For $T<T_{c}$ the expectation value of the Polyakov loop vanishes, and we take $n=0$. Thus, in the latter case
$L_{L}={\rm Re}(L)$ and $L_{T}={\rm Im}(L)$.

\begin{figure}[t]
 \includegraphics[width=3.355in]{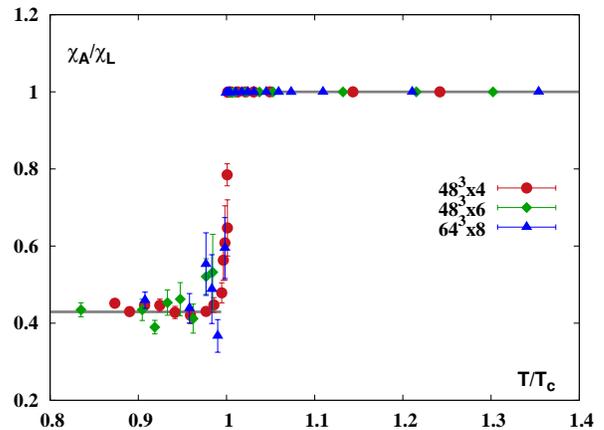}
\caption{The ratio of the modulus $\chi_{A}$  and longitudinal $\chi_{L}$ Polyakov loop susceptibilities obtained in $SU(3)$ lattice gauge theory.
The lines shows $\chi_{A}/\chi_{L}=1$ for $T>T_{c}$ and $\chi_{A}/\chi_{L}=2-\pi/2$ for $T<T_{c}$ (see text).}
\label{ratioA}
\end{figure}

\begin{figure}[t]
\includegraphics[width=3.355in]{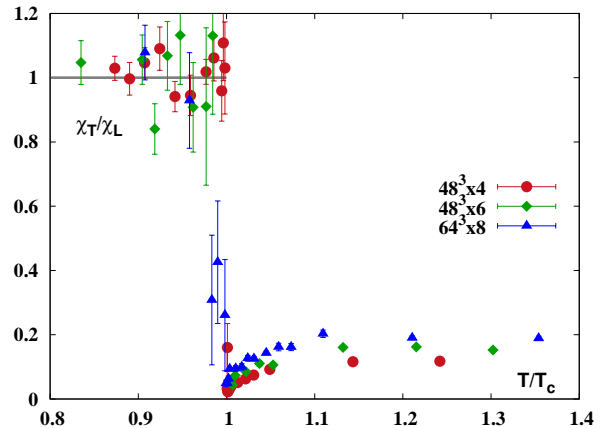}
\caption{
The ratio of the transverse $\chi_{T}$  and longitudinal $\chi_{L}$  Polyakov loop susceptibilities.
The line shows $\chi_{T}/\chi_{L}=1$ for $T<T_{c}$ (see text). }
    \label{ratioI}
\end{figure}

We  have computed  the Polyakov loop susceptibilities, Eqs. \eqref{eq:chiA},
\eqref{eq:chiR} and \eqref{eq:chiI}, within  $SU(3)$ lattice gauge theory,
using the (1,2)-tree-level  Symanzik  improved gauge action on a $N_\sigma^3\times
N_\tau$  lattice. 
We consider lattices of temporal size $N_\tau=4,6$ and $8$ and
spatial extent  $N_\sigma$ varying from 16 to 64. However, in order to make the figures 
more transparent, we show results only for the largest volumes, i.e. for 
$N_{\sigma}= 48$ and 64. We set the temperature for the three temporal lattice extents by varying the bare coupling and use the
temperature scale determined by the zero temperature string tension, as well as the critical couplings of the
deconfinement transition \cite{ref3,Beinlich:1997ia}. The gauge field
configurations were generated using one heatbath and four overrelaxation
updates per sweep with $15\,000$ sweeps in general and up to $100\,000$ sweeps
close to the critical temperature, $T_c$. 

\vskip 0.2cm
\noindent
{{\it The ratios of susceptibilities.}}
In  Figs.~\ref{ratioA} and  \ref{ratioI} we show $SU(3)$ lattice gauge theory results for the  ratios  $R_{A}=\chi_{A}/\chi_{L}$ and $R_{T}=\chi_{T}/\chi_{L}$, 
as functions of temperature. Since renormalization as well as volume and temperature factors cancel in these ratios, they provide robust probes of the 
deconfinement transition. Indeed, both ratios exhibit a strong discontinuity at the deconfinement phase transition and are
almost temperature and volume independent. 
A straightforward interpretation of the properties of  $R_{A}$ and $R_{T}$ is obtained by using   general considerations and the  $\mathcal{Z}_3$ center symmetry.

Consider first the ratio $R_{T}$  for $T<T_c$.  In the $\mathcal{Z}_3$ symmetric phase, the expectation
value of any symmetry breaking operator, e.g. $\tilde{L}$ or $\tilde{L}^{2}$, must vanish. Hence,
\begin{eqnarray}
V(\langle \tilde{L}^2\rangle -\langle \tilde{L}\rangle^2)=\chi_{L}-\chi_{T}=0,
\end{eqnarray}
which implies that   $\chi_{L}=\chi_{T}$. Since $\chi_{L}$ and $\chi_{T}$  are both non-zero,  it follows that  $R_{T}=1$, as shown in Fig.~\ref{ratioI}.

The fact that $R_{A}\simeq 1$ in the deconfined phase, as shown in Fig.~\ref{ratioA}, follows from the following argument.
In the broken symmetry phase, we  introduce shifted operators  $\delta L_{L}$ and $\delta L_{T}$:
\begin{eqnarray}\label{expan}
	L_{L} &= & L_0 + \delta L_{L}, \\
	L_{T} &=& \delta L_{T},
\end{eqnarray}
where $L_0=\langle \tilde{L}\rangle= \langle L_{L}\rangle$ is the (real) expectation val\-ue of the transformed
Polyakov loop and the shifted operators  $\delta L_i$, with $i=(L, T)$,
describe the fluctuations about the mean.
The thermal average of the shifted operator squared  yields the
corresponding susceptibility, $V\langle (\delta L_i)^2\rangle=\chi_{_i}$, $i=L,T$.
We then expand the modulus of the Polyakov loop $|L|=|\tilde{L}|$ in the shifted operators,
\begin{eqnarray}\
	|L| &=& \sqrt{L_{L}^2 + L_{T}^2} \nonumber\\
	    	    &\approx& L_0 \, ( 1 + \frac{\delta L_{L}}{L_0} + \frac{1}{2} \frac{(\delta L_{T})^2}{L_0^2} ).
\label{eq:L-abs}
\end{eqnarray}


Using the fact that by definition $\langle \delta L_{i}\rangle=0$, we find
\begin{eqnarray}\label{r22}
\langle \vert L^{}\vert\rangle &\simeq & L_0 \, ( 1 + \frac{1}{2} \frac{\langle (\delta L_{T})^2\rangle}{L_0^2})
 \end{eqnarray}
 while
 \begin{eqnarray}
\langle \vert L^{}\vert^2\rangle & = & L_0^2 \, ( 1 + \frac{\langle(\delta L_{T})^2\rangle}{L_0^2}+\frac{\langle(\delta L_{L})^2\rangle}{L_0^2})\label{meanl}.
\end{eqnarray}
This implies that to leading order in the expansion of $|L|$, $\chi_{A}\simeq \chi_{L}$ and hence $R_{A}\simeq 1$ for $T>T_{c}$, as shown in Fig.~\ref{ratioA}.

The properties of $R_{T}$ in the  deconfined and $R_{A}$ in the confined phase  cannot be directly linked to the center symmetry.
Using Eq.~\eqref{r22}, we find that
\begin{eqnarray}\label{r2}
\chi_{T}\simeq V(\langle |L|\rangle^2 -\langle L_{L}\rangle^2).
\end{eqnarray}
Thus, in general,  $\chi_{T}$ can be non-vanishing in the $\mathcal{Z}_{3}$  broken phase. However, its value in the high-temperature phase is not constrained by symmetries or general principles. 

In Fig.~\ref{ratioI} we show, that above the phase transition,  $\chi_{T}$ is  in fact much smaller  than $\chi_{L}$.
In the temperature range considered, we find that,  for $T>T_{c}$, the ratio $R_{T}$  is weakly dependent on the  temperature and does not exceed  $\simeq 0.2$. It has been 
argued \cite{Dumitru:2003hp,Dumitru:2002cf} that in the broken $\mathcal {Z}_3$ symmetry phase of the $SU(3)$ gauge theory, $R_{T}$ can be as large as 0.4. We note that in our results, a dependence of $R_{T}$  on $N_\tau$ remains. Hence, we cannot at present draw firm conclusions on the continuum extrapolation of this quantity.

Finally, we turn to the value of $R_{A}$ in the confined phase. In Fig.~\ref{ratioA} we show that, for $T<T_{c}$,  $R_{A}$  is approximately temperature independent, with the lattice results clustering around a value slightly larger than 0.4.
This property of $R_{A}$  can be understood  by  assuming  that in the symmetric phase, the probability distribution for the Polyakov loop is, to a good approximation, Gaussian,
with the  partition function \footnote{More precisely, the quadratic terms of the effective action are responsible for the dominant contribution to the
Polyakov loop susceptibility. However, higher order, non-Gaussian terms are decisive for the determination of
higher order cumulants.}
\begin{eqnarray}\label{gaus}
	Z &= \int \, d L_{L} d L_{T} \,  e^{- V T^3 [\alpha(T) (L_{L}^2 + L_{T}^2)]},
\end{eqnarray}
where the integrations extend from $-\infty$ to $\infty$.
The susceptibilities are then obtained by performing elementary integrals
\begin{eqnarray}
\chi_{L}  &=& \frac{1}{2 \alpha\,T^{3}},~~
	\chi_{T}  =\frac{1}{2 \alpha\,T^{3}},~~\nonumber\\
	\chi_{A} &=& \frac{1}{2 \alpha\,T^{3}} \left(2-\frac{\pi}{2}\right).
\end{eqnarray}
Consequently,
$R_A=(2-\pi/2)\simeq 0.429$, in good agreement  with the  lattice results, shown
in Fig.~\ref{ratioA}. We note that the Gaussian approximation is not expected to be valid
close to $T_{c}$, where the coefficient $\alpha(T)$ in (\ref{gaus}) is small and hence  higher order terms cannot be neglected.

In $SU(2)$ gauge theory the Polyakov loop is real, so the corresponding integrals are one-dimensional,
which implies a slightly different ratio, $R_A^{SU(2)}=(1-2/\pi)\simeq 0.363$. This value is indeed in agreement with lattice results for the $SU(2)$ Polyakov loop susceptibilities below $T_{c}$, outside of the critical region~\cite{Engels:1998nv}.
These results indicate that in the symmetric phase, the effective Polyakov loop potential is well approximated by a Gaussian both in $SU(2)$ and $SU(3)$ lattice gauge theories.

\vskip 0.2cm
\noindent
{\it Conclusions.}
We have shown, that the
ratios of  Polyakov loop susceptibilities provide an excellent signal for  the deconfinement
phase transition in $SU(3)$ gauge theory. The ratios  are discontinuous  at the phase transition and only weakly  temperature dependent on either side
of  $T_c$. Moreover, they are independent of the Polyakov loop renormalization and only weakly dependent on the system size.

We have also shown that, with one exception, the ratios obtained  outside of the transition region,  can be understood in
terms of  general symmetry arguments and
the observation, that in the confined phase, the Polyakov  loop probability distribution is well approximated by a
Gaussian. There is, however, no restriction by  symmetry on the ratio of the transverse to longitudinal susceptibility, $\chi_{T}/\chi_{L}$ in the deconfined phase.
We find that, above $T_{c}$ this ratio  is fairly small and varies weakly with  temperature.

In  QCD, the global $\mathcal{Z}_3$ symmetry  is explicitly broken by finite quark masses. Hence, the properties of the susceptibility ratios in QCD can
differ from those in pure gauge
theory. In particular, the discontinuity will most likely be smoothened, since in QCD the transition is continuous.
Nevertheless, outside of the transition region, the ratios may approximately reflect the constraints from center symmetry and
thus provide a useful probe of the confinement-deconfinement transition also in full QCD.

\acknowledgements
We acknowledge stimulating discussions with   Frithjof  Karsch.
P.M.Lo acknowledges the support of the Frankfurt Institute for Advanced Studies (FIAS).
B.F. is supported in part by the Extreme Matter Institute EMMI.
K.R. acknowledges partial support of the Polish Ministry of National
Education (NCN). The work of C.S. has been partly supported
by the Hessian LOEWE initiative through the Helmholtz
International Center for FAIR (HIC for FAIR). The numerical calculations
have been performed on the Bielefeld GPU Cluster.


\begin{thebibliography}{10}

\bibitem{sy1}
 B.~Svetitsky,  and L.~G.~Yaffe,
  Nucl.\ Phys.\ B {\bf 210} (1982) 423.

\bibitem{sy2}
   L.~G.~Yaffe,  and B.~Svetitsky,
  Phys.\ Rev.\ D {\bf 26} (1982) 963.



\bibitem{Boyd:1995zg}
 G.~Boyd, J.~Engels, F.~Karsch, E.~Laermann, C.~Legeland, M.~Lutgemeier,  and B.~Petersson,
  Phys.\ Rev.\ Lett.\  {\bf 75} (1995) 4169

\bibitem{su31}
G.~Boyd, J.~Engels, F.~Karsch, E.~Laermann, C.~Legeland, M.~Lutgemeier, and B.~Petersson,
  Nucl.\ Phys.\ B {\bf 469} (1996) 419.



\bibitem{ref1}
M. Fukugita, M. Okawa, and A. Ukawa, Nucl. Phys.
{ B} {\bf 337} (1990) 181.

\bibitem{Borsanyi:2012ve}
S.~Borsanyi, G.~Endrodi, Z.~Fodor, S.~D.~Katz,  and K.~K.~Szabo,
  JHEP {\bf 1207} (2012) 056.

 \bibitem{McLerran:1980pk}
  L.~D.~McLerran,  and B.~Svetitsky,
  Phys.\ Lett.\ B {\bf 98} (1981) 195.
   L.~D.~McLerran,  and B.~Svetitsky,
  Phys.\ Rev.\ D {\bf 24} (1981) 450.

\bibitem{general}
 A. M. Polyakov, Phys. Lett.B {\bf 72} (1978) 477. G.'t Hooft,
Nucl. Phys. B {\bf 138} (1978) 1. L. Susskind, Phys.Rev.D {\bf 20} 
 (1979) 2610.

\bibitem{ref2}
Y. Iwasaki
{\it et al.}, Phys. Rev. D
{\bf 46}  (1992) 4657.
\bibitem{ref3}
R. G. Edwards, U. M. Heller, and T. R. Klassen, Nucl. Phys.
{B} {\bf 517} (1998) 377.
\bibitem{ref4}
G. Boyd, J. Engels, F. Karsch, E. Laermann, C. Legeland, M.
Lutgemeier, and B. Petersson, Phys. Rev. Lett.
{\bf 75} (1995) 4169;  Nucl. Phys.
{ B} {\bf 469} (1996) 419.
\bibitem{ref5}
T. DeGrand
{\it et al.}, Nucl. Phys.
{B} {\bf 454} (1995) 615.  

\bibitem{Kaczmarek:2002mc}
O.~Kaczmarek, F.~Karsch, P.~Petreczky,  and F.~Zantow,
  Phys.\ Lett.\ B {\bf 543} (2002) 41.
  
\bibitem{gavai} 
U. M. Heller,  and F. Karsch, Nucl. Phys. B
{\bf 251}
(1985) 254.  
 R.~V.~Gavai,
  Phys.\ Lett.\ B {\bf 691} (2010) 146. 
  S.~Gupta, K.~Huebner,  and O.~Kaczmarek,
  Phys.\ Rev.\ D {\bf 77} (2008) 034503.

\bibitem{Beinlich:1997ia}
   B.~Beinlich, F.~Karsch, E.~Laermann and A.~Peikert,
   Eur.\ Phys.\ J.\ C {\bf 6} (1999) 133.
 
  \bibitem{Dumitru:2003hp}
 A.~Dumitru, Y.~Hatta, J.~Lenaghan, K.~Orginos,  and R.~D.~Pisarski,
  Phys.\ Rev.\ D {\bf 70} (2004) 034511.

\bibitem{Dumitru:2002cf}
A.~Dumitru,  and R.~D.~Pisarski,
  Phys.\ Rev.\ D {\bf 66} (2002) 096003. 

\bibitem{Engels:1998nv}
J.~Engels,  and T.~Scheideler,
  Nucl.\ Phys.\ B {\bf 539} (1999) 557.


\end{thebibliography}
\end{document}